\documentclass[aps,prl,reprint,groupedaddress,amsmath,showpacs,nofootinbib]{revtex4-1}
\usepackage{scalefnt}
\usepackage{braket}
\usepackage{xfrac}
\usepackage{amssymb}
\usepackage{graphicx}
\usepackage{xcolor}
\usepackage{isotope}
\usepackage{pifont}
\usepackage{siunitx}
\usepackage{hyperref}
\usepackage{cleveref}
\usepackage{times,txfonts}

\newcommand{\op}[1]{\ensuremath{\hat{#1}}}
\newcommand{\Nmax}{\ensuremath{N_\text{max}}}
\newcommand{\Tmax}{\ensuremath{T_\text{max}}}
\newcommand{\emax}{\ensuremath{e_\text{max}}}
\newcommand{\hw}{\ensuremath{\hbar\Omega}}

\begin{document}

\title{Electromagnetic Strength Distributions from the \emph{Ab Initio} No-Core Shell Model}

\author{Christina Stumpf}
\email[]{christina.stumpf@physik.tu-darmstadt.de}
\author{Tobias Wolfgruber}
\author{Robert Roth}
\affiliation{Institut f\"ur Kernphysik, TU Darmstadt, Schlossgartenstr. 2, 64289 Darmstadt, Germany}

\date{\today}

\begin{abstract}

We present an \emph{ab initio} approach for the description of collective excitations and transition strength distributions of arbitrary nuclei up into the  sd-shell that based on the No-Core Shell Model in combination with the Lanczos strength-function method. Starting from two- and three-nucleon interactions from chiral effective field theory, we investigate the electric monopole, dipole, and quadrupole response of the even oxygen isotopes from \isotope[16]{O} to \isotope[24]{O}. The method describes the full energy range from low-lying excitations to the giant resonance region and beyond in a unified and consistent framework, including a complete description of fragmentation and fine-structure. This opens unique opportunities for understanding dynamic properties of nuclei from first principles and to further constrain nuclear interactions. We demonstrate the computational efficiency and the robust model-space convergence of our approach and compare to established approximate methods, such as the Random Phase Approximation, shedding new light on their deficiencies. 
 
\end{abstract}


\maketitle


\paragraph{Introduction.}

The theoretical description of atomic nuclei based on \emph{ab initio} solutions of the nuclear many-body problem has become a very dynamic and productive field of research, driven by two developments: (i) the construction of consistent and systematically improvable nuclear interactions in chiral effective field theory (EFT) \cite{Epelbaum2008,EpHa09,Machleidt2011}, and (ii) the formulation of new and refined approaches for the solution of the nuclear many-body problem with controlled uncertainties. 
Most of the applications of \emph{ab initio} approaches are restricted to ground and low-lying excited states and their properties \cite{Hjorth-Jensen2017}.

Collective excitations of nuclei are largely uncharted territory from the point of view of \emph{ab initio} nuclear structure theory. The study of collective modes, e.g., the giant electric monopole, dipole, and quadrupole resonances, as well as the electromagnetic and weak response in general, has a long history in nuclear structure physics \cite{Speth1991,Harakeh2001,Rowe2010}. Many new aspects, e.g., the so-called pygmy dipole resonance \cite{Savran2013}, the fragmentation and fine structure of giant resonances \cite{NeumannCosel1999,Lacroix2000,Shevchenko2004,Kalmykov2006,Shevchenko2008,Usman2011,Poltoratska2014}, the response of neutron-rich nuclei \cite{Leistenschneider2001,Klimkiewicz2007}, and the connection of dynamic and static properties \cite{Reinhard2010,Piekarewicz2011,Tamii2011,Piekarewicz2012,Roca-Maza2015,Birkhan2017}, are at the heart of ongoing experimental and theoretical investigations. 
Collective excitations serve as a magnifying glass for the internal dynamics of the nucleus and provide additional and complementary information on the effects of nuclear interactions. Knowledge of the nuclear response to electromagnetic probes is important for many applications, e.g., for reaction mechanisms involved in nucleosynthesis processes, such as the $p$- and $r$-process \cite{Arnould2003,Arnould2007}. 

On the theory side, the description of collective excitations is still dominated by phenomenological models. In the domain of microscopic theories based on energy-density functionals or phenomenological mean-field interactions, the Random Phase Approximation (RPA) \cite{Ring1980} and its extensions are the main workhorse. It is clear from the outset that simple RPA, built from one-particle-one-hole excitations, will not suffice to describe the rich physics of the nuclear response and extensions, such as Second RPA (SRPA) including two-particle-two-hole excitations \cite{DaProvidencia1965,Yannouleas1987,Drozdz1990}, will be necessary.
Applications of RPA-type methods with realistic nuclear interactions underline the importance of including ground-state correlations in these approaches \cite{Papakonstantinou2010}. These limitations demonstrate the need for a complete \emph{ab initio} description of the transition strength distribution of nuclei.  

First steps to describe collective excitations in an \emph{ab initio} framework beyond light nuclei have been made using the Lorentz-integral transform (LIT) combined with the coupled-cluster method \cite{Bacca2013,Bacca2014,Hagen2015,Miorelli2016,Birkhan2017}. The LIT maps the continuum problem onto a bound-state-problem and, in this way, avoids the challenging calculation of final states in the continuum. However, the derivation of response functions requires a delicate inversion procedure, which only yields the gross structure of giant resonances above the particle threshold. 


In this Letter, we present a new approach for the \emph{ab initio} description of collective excitations and the nuclear response.
We combine the No-Core Shell Model (NCSM) \cite{Navratil2009,Barrett2013} with the Lanczos strength-function method proposed by Whitehead \cite{Whitehead1980} for the efficient computation of transition strengths and their distribution. This enables direct \emph{ab initio} calculations of all relevant strength distributions for all nuclei up into the lower sd-shell. In addition, we can test the validity of established approximate methods, like RPA and SRPA, and elucidate some of their inherent deficiencies.

\paragraph{Method.}

Our \emph{ab initio} scheme for the calculation of strength distributions for specific electromagnetic transition operators involves two steps.  
First, we solve the eigenvalue problem of the Hamiltonian using the NCSM within a model space truncated with respect to the maximum number of excitation quanta $\Nmax$. Since the basis dimensions grow factorially with $A$ and $\Nmax$, we additionally employ the importance truncation \cite{Roth2007,Roth2009}. The ground-state eigenvector $\ket{\Psi_0}$ serves as input for the next step of the calculation. 

In a second step, following Whitehead \cite{Whitehead1980}, we iteratively construct fast converging approximations of the strength distribution using the Lanczos method \cite{Lanczos1950,Paige1971} in its simplest form.
This method has been used successfully for the calculation of electromagnetic or weak responses in the shell model for some time \cite{Caurier2005,
Caurier1990,Engel1992,Caurier1994,Caurier1995,Gueorguiev2000,Stetcu2003,Haxton2005,Loens2012,Shimizu2017}.
We construct a normalized pivot vector $\ket{v_1}$ by applying the transition operator $\op{O}_\lambda$, e.g., an electromagnetic multipole operator, to the ground-state eigenvector $\ket{\Psi_0}$ obtained in the first step:
\begin{align} \label{eq:pivot}
 \ket{v_1} = \frac{1}{\sqrt{S}} \op{O}_\lambda \ket{\Psi_0}.
\end{align}
The normalization factor, $S = \braket{\Psi_0 | \op{O}_\lambda^\dagger \op{O}_\lambda | \Psi_0}$, corresponds to the total transition strength from the ground state $\ket{\Psi_0}$ to any excited state.
Starting from $\ket{v_1}$, we use the simple Lanczos algorithm with the Hamiltonian to iteratively construct an orthonormal Lanczos basis $\{ \ket{v_i} \}$ in which the Hamilton matrix is tridiagonal. We carry out $p$ iterations to construct the tridiagonal $p\times p$ matrix $T$ and obtain its eigenvalues $E_n$ and eigenvectors $C_{n,i}$. The eigenvectors define approximations for $p$ eigenstates of the Hamiltonian via $\ket{E_n} = \sum_{i=1}^p C_{n,i} \ket{v_i}$. In standard applications, we would continue the iterations until the eigenvalues and eigenstates of interest are converged. 

For evaluating transition strengths, the first coefficient $C_{n,1}$ in each eigenvector of the $T$ matrix plays an important role.  
Using $C_{n,1} = \braket{ E_n | v_1 }$ and the special definition of the pivot $\ket{v_1}$ given in (\ref{eq:pivot}), the reduced transition matrix element between a $J=0$ ground state $\ket{\Psi_0}$ and the Lanczos approximation for an excited state $\ket{E_n}$ is given by $ | \braket{ E_n || \op{O}_\lambda || \Psi_0 } |^2 = (2 \lambda 
+ 1) S |C_{n,1}|^2$. With this, we construct the discrete strength distribution
\begin{align}
 R(E^*) &= \sum_{n} | \braket{ E_n || \op{O}_\lambda || \Psi_0} |^2\; \delta{(E^*-(E_n - E_0))}. \label{eq:strengthfunction}
\end{align}
The fact that we obtain a discrete excitation spectrum results from the use of a bound-state method, the NCSM. The coupling to the continuum and the resulting escape width are not captured, however, all correlation effects are explicitly taken into account. In practical applications, we are often interested in smoothed-out distributions, lending themselves to an easy comparison with experimental data. Therefore, we mainly discuss continuous strength functions obtained by folding $R(E^*)$ with a Lorentzian of 1 MeV width.
 
A brief comment on computational aspects: The calculation of strength distributions does not entail additional computational limitations. Generally, if the NCSM calculation for the ground state is feasible, then we can also compute the strength distribution. One reason is that a few hundred Lanczos iterations are generally sufficient to fully converge the strength functions. Furthermore, we only need to store three Lanczos vectors, since we work with the simplest possible version of the algorithm. A disadvantage of the simple Lanczos algorithm is the loss of orthogonality of the Lanczos basis due to round-off errors for finite machine precision. However, the resulting duplicates in the energy spectrum do not affect the strength distributions, as \eqref{eq:strengthfunction} implies a summation of strength corresponding to the same energy. We have confirmed that strength functions obtained in this way are identical to those obtained in calculations with an explicit reorthogonalization of the Lanczos basis.

\paragraph{Calculation Details.}
In the following calculations, we employ two nuclear Hamiltonians with chiral nucleon-nucleon (NN) and three-nucleon (3N) interactions.
The interaction denoted by NN+3N(400) consists of an NN interaction at next-to-next-to-next-to leading order by Entem and Machleidt \cite{Entem2003} and a 3N interaction at next-to-next-to leading order with a local regulator and cutoff $\Lambda_{3N} = 400 \, \text{MeV}$ \cite{Navratil2007,Roth2012}.
We further use the NNLO$_\text{sat}$ \cite{Ekstrom2015} interaction, where the NN and 3N force is derived consistently up to next-to-next-to leading order and reproduces the empirical saturation properties.
Both Hamiltonians are softened using a Similarity Renormalization Group (SRG) transformation with flow parameter $\alpha = 0.08 \, \text{fm}^4$ \cite{RoLa11,BoFu07,HeRo07,RoRe08,Jurgenson2009,Jurgenson2013,Roth2014}. Induced many-nucleon forces are included consistently up to the 3N level, contributions beyond that level are neglected.
To limit the computational cost, we employ the normal-ordered two-body approximation \cite{Roth2012}.
We consider electric multipole transitions mediated by the standard operators---not SRG-evolved---in isospin-decomposed form \cite{Harakeh2001}.
In the NCSM, we use a Hartree-Fock basis with single-particle truncation $\emax=12$ in order to improve convergence and reduce frequency dependencies.

\begin{figure*}
\includegraphics[width=\textwidth]{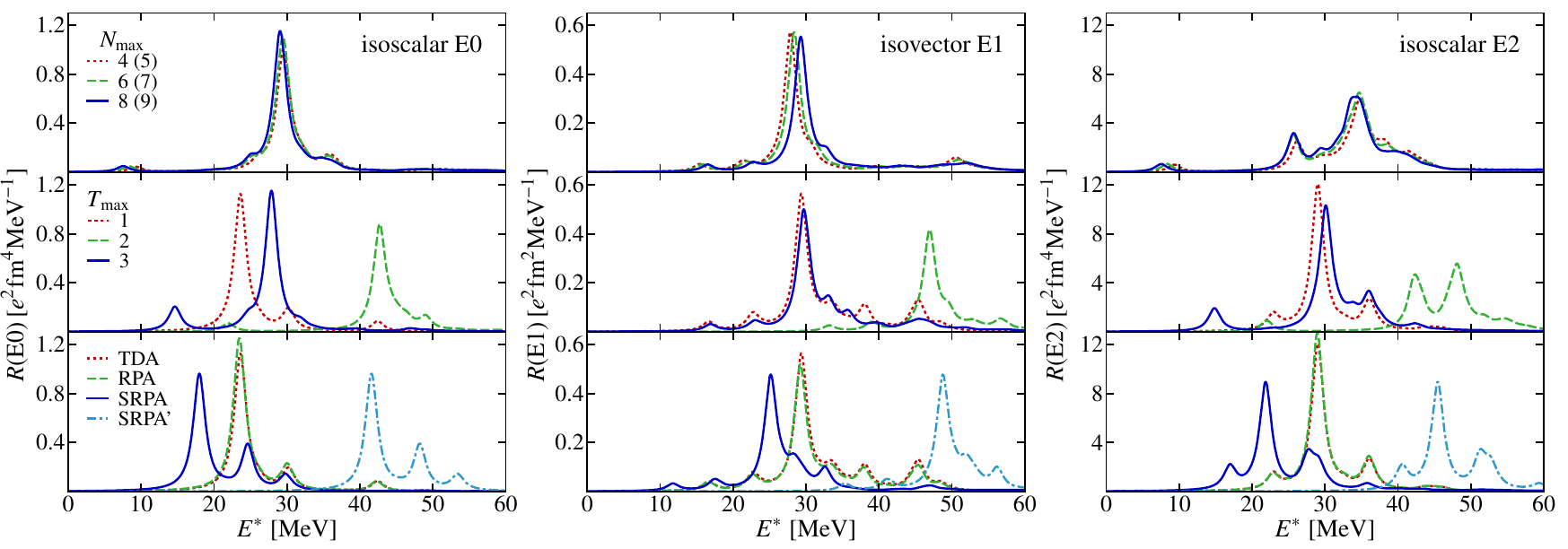}
\caption{(color online) Isoscalar E0, isovector E1, and isoscalar E2 strength functions of \isotope[16]{O} using the SRG-evolved NN+3N(400) interaction and $\hw=24 \, \text{MeV}$.
The rows show the convergence behavior of strength functions obtained in the NCSM with $\Nmax$ truncation, NCSM with $\Tmax$ truncation, and the TDA, RPA, and SRPA.
SRPA' refers to the SRPA strength shifted by the second-order perturbative energy correction. \label{fig1}
}
\end{figure*}

\paragraph{Convergence and Validation.}
We begin with studying the convergence behavior of the isoscalar electric monopole (E0), isovector electric dipole (E1), and isoscalar electric quadrupole (E2) strength distributions for \isotope[16]{O} using the NN+3N(400) interaction. The top row in Fig.~\ref{fig1} illustrates the convergence of the continuous strength functions with increasing $N_{\max}$. Already for moderate model-space sizes, the results are remarkably stable, both, for the prominent giant resonance features and for the smaller structures. The largest systematic dependence of $\Nmax$ appears for the isovector E1 transitions that connect the $J^{\pi}=0^+$ ground state to $1^-$ unnatural parity states, which are known to exhibit slowly converging excitation energies in the NCSM.
The low-lying strength around 7-8 MeV in the isoscalar E0 and E2 strength functions is due to center-of-mass contaminations and vanishes if a Lawson-type center-of-mass term $\beta \hat{H}_\text{cm}$ \cite{Gloeckner1974313} is added, as we will do in the later applications. We note that the effect of the importance truncation (not shown) is very small and less significant than the model-space truncations. Analogous convergence studies for the number of Lanczos iterations show that the strength distributions are completely converged after about 100 iterations---throughout this Letter we show results obtained for at least 400 Lanczos iterations. 

An interesting alternative to the $\Nmax$ truncation of the many-body basis is a particle-hole or $\Tmax$ truncation, where $\Tmax$ denotes the maximum number of particles promoted to excited single-particle orbits. The middle row in Fig.~\ref{fig1} shows the dependence of the strength distributions on $\Tmax$ in an $\emax=12$ single-particle space. The effects are quite dramatic: Whereas the giant resonances for $\Tmax=1$, i.e., a one-particle-one-hole (1p1h) space, appear at reasonable energies, the $\Tmax=2$ strength is shifted by about 20 MeV to higher energies. For $\Tmax=3$ the resonance energies are shifted back into the neighborhood of their original position. The strength distribution for $\Tmax=3$ agrees very well with the converged result for $\Nmax$-truncated spaces for the E0 and E1 modes. For E2 some fragmentation is still missing, which appears only after including $\Tmax=4$ configurations (not shown). The strong impact of $\Tmax=3$ configurations was also found in shell-model calculations for the Gamow-Teller strength distribution \cite{Caurier1994}.
This behavior can be explained quite intuitively: Since we start from a HF basis and by means of Brillouin's theorem, the HF ground state does not couple to one-particle one-hole configurations. Thus, for $\Tmax=1$ only the excited states are built from 1p1h excitations. Including 2p2h configurations for $\Tmax=2$ into the model space causes a large shift of the ground state energy by about $-25\,\text{MeV}$, but smaller shifts of about $-5\,\text{MeV}$ on the absolute energies of the excited states. Therefore, resonances are shifted to unrealistically large excitation energies, which is clearly an artifact of the $\Tmax=2$ truncation. At $\Tmax=3$ this problem is remedied and both, ground state and 1p1h dominated excitations, acquire the important 2p2h corrections such that excitation energies move back to realistic values. Note that this imbalance of correlation content in ground and excited states does not appear in $\Nmax$-truncated model spaces. 

The $\Tmax$ truncation provides a natural link to traditional RPA-type methods. In the bottom row of Fig.~\ref{fig1}, we present results of conventional RPA and SRPA calculations for that same interaction and single-particle space as the $\Tmax$-truncated NCSM calculations. The $\Tmax=1$ calculations are formally equivalent to the Tamm-Dancoff Approximation (TDA), which can be viewed as a simplification of RPA omitting the backward amplitudes associated with hole-particle terms. Also numerically, the results agree perfectly. Compared to TDA, the strengths obtained from a self-consistent RPA calculation differ only slightly, i.e., the impact of the backward amplitudes is small. 
Going from RPA to SRPA, i.e., including 2p2h degrees of freedom into the excitation operator, shifts the strength to lower energies, as illustrated in Fig.~\ref{fig1}. This shift is a well-known problem of SRPA, which has received quite some attention in the recent literature \cite{Papakonstantinou2009,Papakonstantinou2010,Gambacurta2010,Gambacurta2011,Gambacurta2011a,Gambacurta2012,Tselyaev2013,Papakonstantinou2014,Gambacurta2015,Gambacurta2016a,Gambacurta2017}. Apart from additional issues with double counting in density-functional based calculations \cite{Tselyaev2013,Gambacurta2015}, this shift is associated with an inconsistency of using the HF ground state when constructing the SRPA equations that include explicit 2p2h excitations \cite{Papakonstantinou2010,Papakonstantinou2014}. Effectively, a standard SRPA calculation yields the energies of the excited modes relative to the HF ground-state energy and not relative to a correlated ground state including 2p2h configurations. We can mimic the effect of 2p2h ground-state correlations by shifting the SRPA excitation energies by the second-order perturbative correction to the ground-state energy, which is a simple means to quantify the effect of 2p2h admixtures to the ground state. The resulting strength distribution is denoted SRPA' in Fig.~\ref{fig1} and agrees well with the $\Tmax=2$-truncated NCSM result, as expected on the basis of the particle-hole content of the model space. Then, however, SRPA' suffers from the same problem as the $\Tmax=2$ model space that we discussed above. This comparison provides a different perspective on the consistency issues of SRPA and helps to devise approaches to resolve them \cite{Trippel2016}.  

\begin{figure}
 \includegraphics[width=1\columnwidth]{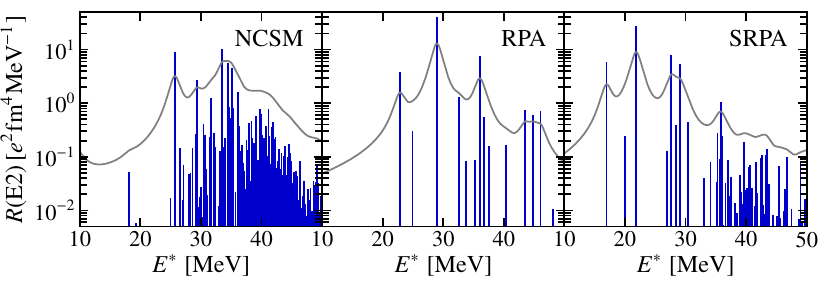}
\caption{(color online) Discrete isoscalar E2 strength distribution for \isotope[16]{O} obtained in the NCSM for $\Nmax=8$, RPA, and SRPA using the SRG-evolved NN+3N(400) interaction and $\hw=24\,\text{MeV}$. \label{fig2}
}
\end{figure}

\begin{figure*}
 \includegraphics[width=\textwidth]{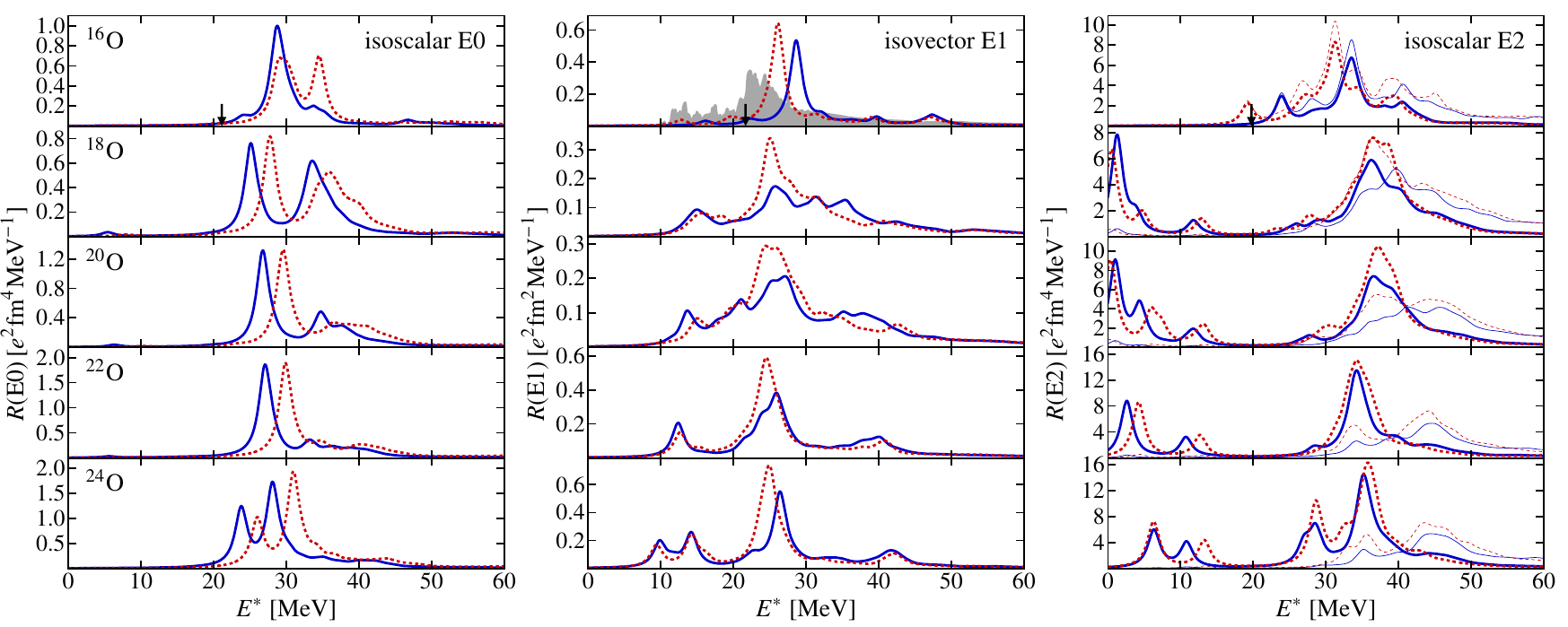}
\caption{(color online) Isoscalar E0, isovector E1, and isoscalar E2 strength functions of the even-$A$ oxygen isotopes \isotope[16-24]{O} using the SRG-evolved NN+3N(400) (blue line) and NNLO$_\text{sat}$ (red dotted line) interaction. The thin lines in the right-hand column show the total E2 response.
The oscillator frequency is $\hw=20\,\text{MeV}$ and $\beta=0.5$.
The NCSM model spaces are truncated at $\Nmax=8(9)$.
The arrows indicate the experimental centroid energies for \isotope[16]{O} from \cite{Lui2001}, the gray area shows experimental data from \cite{Ahrens1975} in arbitrary units. \label{fig3}
}
\end{figure*}

One of the main motivations to use SRPA is the description of fragmentation and fine structure of resonances, which is observed in experiment \cite{Lacroix2000,Shevchenko2004,Shevchenko2008,Savran2008,Tonchev2010,Usman2011,Poltoratska2014}. We will not elaborate on this interesting topic, but rather show that the strength distributions from the NCSM predict substantial fine structure. In Fig.~\ref{fig2}, we compare the discrete strength distributions from NCSM, RPA, and SRPA on a logarithmic scale for the example of the isoscalar E2 mode in \isotope[16]{O}. Evidently, the NCSM strength shows much more fragmentation and fine structure in the energy region of the giant resonance than even SRPA.

\paragraph{Applications.}

A particularly interesting region for applications of the NCSM for strength distributions is the oxygen isotopic chain. It is easily within the reach of the method and the collective response of the neutron-rich oxygen isotopes has been and continues to be a focus of research \cite{Sagawa1999,Leistenschneider2001}. In Fig.~\ref{fig3} we present the isoscalar E0, isovector E1, and isoscalar E2 strength distributions for the even oxygen isotopes from \isotope[16]{O} to \isotope[24]{O} for the NN+3N(400) and the NNLO$_{\text{sat}}$ interaction. An important difference between the two chiral NN+3N interactions is that NN+3N(400) underestimates the ground-state radii of the oxygen isotopes by about 10\%, while NNLO$_{\text{sat}}$ is constructed to reproduce the experimental radii well.

Generally, the structure of the response is very similar for the two interactions. For the isoscalar E0 strength the main difference is a relative shift of the stength distribution, with the NNLO$_{\text{sat}}$ interaction producing about 4 MeV higher resonance energies. This is surprising since in a naive mean-field picture one would expect lower resonance energies for an interaction that produces larger ground-state radii. The comparison of the NNLO$_{\text{sat}}$ response for \isotope[16]{O} to the experimental centroid energy for the isoscalar giant monopole resonance is also surprising. Our predicted monopole resonance appears at too high energies although the interaction is known to predict a nuclear matter incompressibility within the empirical range \cite{Ekstrom2015,Stone2014}. This indicates that other aspects of the interaction, e.g., momentum dependencies or non-localities, play an important role for the transition strength that is not probed by static properties. The overestimation of the resonance energies compared to experiment is also evident for the E1 and E2 strength distributions, although the NNLO$_{\text{sat}}$ interaction tends to predict a lower resonance energy than the NN+3N(400) interaction for these modes. 

The strength distributions exhibit interesting systematics throughout the isotopic chain. The isovector E1 distribution broadens as one moves toward mid-shell at \isotope[20]{O}, as observed experimentally \cite{Leistenschneider2001}, and narrows again as the next neutron closed-shell is approached.
At the same time more and more low-energy strength appears, which is compatible with the emergence of pygmy dipole excitations  \cite{Chambers1994,Suzuki1990,Savran2013}.
We can calculate electric dipole polarizabilities from the strength functions, which, however, are too small because the strength is at too high energies.
For the NNLO$_\text{SAT}$ interaction, we obtain a dipole polarizability of $0.48\,\text{fm}^3$ compared to the experimental value $0.58\,\text{fm}^3$ \cite{Ahrens1975}.
With adding more and more neutrons, our predictions for the dipole polarizabilities increase systematically, reaching $1.09\,\text{fm}^3$ for \isotope[24]{O}. The isoscalar E2 distribution starting from \isotope[18]{O} shows strong contributions from low-lying neutron-dominated $2^+$ excitations, which hardly contribute to the total E2 strength without isospin decomposition, also shown in Fig.~\ref{fig3}. This low-lying quadrupole strength can be interpreted as a pygmy quadrupole resonance, predicted in \cite{Tsoneva2011} and recently measured for \isotope[124]{Sn} \cite{Spieker2016}.

\paragraph{Conclusions.}
We have formulated an \emph{ab initio} approach for the description of transition strength distributions by combining the NCSM with the Lanczos strength-function method. It solely relies on a truncation of the many-body basis and we demonstrated convergence of the strength distributions with the truncation parameter $\Nmax$. Our approach provides access to the low-energy strength and to the giant resonance region including fragmentation and fine structure. Only the explicit coupling to the continuum, e.g., to describe the escape width above the relevant particle threshold, is not included. In this respect our method is complementary to LIT approaches, which formally include continuum physics but cannot address the (sub-) threshold region as well as fragmentation and fine structure. We elaborated on the relation with RPA-type approaches and shed new light on the deficiencies of the SRPA method. Our applications to neutron-rich oxygen isotopes highlight the significance of collective modes for constraining nuclear interactions.

\paragraph{Acknowledgments.}
This work is supported by the BMBF through contracts 05P15RDFN1 (NuSTAR.DA) and 05P2015 (NuSTAR R\&D), the DFG through contract SFB 1245, and the Helmholtz International Center for FAIR. Numerical calculations have been performed at the computing center of the TU Darmstadt (Lichtenberg) and at the LOEWE-CSC Frankfurt.

%

\end{document}